\documentstyle[aps,twocolumn,floats,prl]{revtex}
\input epsf

\newcommand\lsim{\mathrel{\rlap{\lower4pt\hbox{\hskip1pt$\sim$}}
        \raise1pt\hbox{$<$}}}
\newcommand\gsim{\mathrel{\rlap{\lower4pt\hbox{\hskip1pt$\sim$}}
        \raise1pt\hbox{$>$}}}

\begin{document}

\twocolumn[\hsize\textwidth\columnwidth\hsize\csname
@twocolumnfalse\endcsname

\title{Short Distance Physics and the Consistency Relation
for Scalar and Tensor Fluctuations in the Inflationary Universe}

\author{Lam Hui and William H. Kinney
}

\address{
Department of Physics,
Columbia University, 538 West 120th Street, New York, NY 10027\\
Institute for Strings, Cosmology and Astro-particle Physics (ISCAP)\\
{\tt lhui@astro.columbia.edu, kinney@physics.columbia.edu}}

\date{\today}
\maketitle
\begin{abstract}
Recent discussions suggest the possibility that short distance physics can significantly
modify the behavior of quantum fluctuations in the inflationary universe, and alter
the standard large scale structure predictions.
Such modifications can be viewed as due to a different choice of the vacuum state.
We show that such changes generally lead to violations of the well-known consistency
relation between the scalar to tensor ratio and the tensor spectral index.
Vacuum effects can introduce an observable modulation to the usual
predictions for the scalar and tensor power spectra.
\vskip 0.3truecm
98.80.Bp; 98.80.Cq; 98.65.Dx; 04.62.+v
\end{abstract}  
\vskip 0.5truecm
]


\section{Introduction}
\label{intro}

One of the key predictions of inflation \cite{inflation} is a spectrum of scalar
and tensor fluctuations, which are quantum in origin and
are stretched from very small wavelengths to cosmological 
scales \cite{fluc}. Recently, there has been much discussion of whether
short distance physics might modify the nature of the fluctuations.
While our current ignorance about physics at 
very high energy makes a definite prediction difficult,
examples have been put forward where
the standard predictions for the amplitude and spectrum of
fluctuations are modified \cite{transplanck}. 
Most focus on scalar fluctuations
\cite{note}, whose amplitude and spectrum are also determined
by the choice of the inflaton potential in the standard theory.
Modifications in the scalar fluctuations due to
short distance physics might therefore be difficult
to tell apart observationally from those due to a different choice
of the inflaton potential. 

This motivates us to examine the relation between scalar
and tensor fluctuations generated during inflation.
It is well known that for the simplest inflation
models, single-field slow-roll inflation, a certain
consistency relation exists between the ratio of
tensor to scalar amplitude and the tensor spectral index \cite{consistency}.
It has been argued that such a relation, if observationally verified,
would offer strong support for the idea of inflation.
Here, we demonstrate that modifications of
short distance physics generally lead to violations
of this consistency relation (and its generalization
in multiple-field models). This opens up the exciting
possibility of probing new physics at high energies if tensor modes are
detected \cite{gravitywaves}.

To make our discussion general, we choose
not to focus on particular modifications of short distance
physics, but instead view such modifications as equivalent to
a different choice of the vacuum state. Indeed, if
new physics modifies the dynamics of fluctuations only on
some (proper) scales much smaller than the Hubble radius during
inflation, as far as the late-time evolution of 
a given comoving wave-mode is concerned, the only difference
from the standard theory is that the initial condition or
vacuum state has been modified \cite{tanaka}.
The standard choice is known as the Bunch-Davies vacuum,
which corresponds to the Minkowski vacuum in the small
wavelength limit \cite{bunchdavies}. 
This is in some sense the most natural vacuum. For instance,
it respects the symmetry of the de Sitter group. On the other hand,
inflation generally does not correspond to exact de Sitter expansion. 
It is conceivable that
physics at very short distances effectively forces another choice
of the vacuum. Such examples have been constructed, and 
it would be useful to investigate the generic observational
consequences.

Our discussion is organized as follows. In \S \ref{vacuum}, 
we derive the scalar and tensor power spectra, allowing for
a more general vacuum. We then derive the tensor to scalar ratio
and their respective spectral indices
in \S \ref{discuss}, and show how the consistency relation can
be broken.
There is naturally the concern that a different choice of
the vacuum might give rise to a particle background which
contributes significantly to the total energy density,
hence violating the condition for inflation \cite{tanaka}. 
This, and multiple-field inflation, will be discussed in
\S \ref{discuss} as well.

\section{Scalar and Tensor Power Spectra and the Choice of the Vacuum}
\label{vacuum}

Here, we follow closely the notation of \cite{lidsey}. Much of
the material here has been derived before assuming
the Bunch-Davies vacuum. We repeat the essential steps, 
showing where and how the choice of the vacuum affects
the final predictions \cite{huang}.
The metric can be written as 
\begin{eqnarray}
ds^2 &=& a(\tau)^2 [- (1 + 2 {\cal A}) d\tau^2 + 
2 B_{,i} d \tau dx^i + \\ \nonumber 
&& ( (1 - 2\Psi) \delta_{ij} + 2 E_{,ij} + h_{ij})
dx^i dx^j]
\end{eqnarray}
where $a(\tau)$ is the Hubble scale factor as a function of
conformal time $\tau$, ${\cal A}$, $B$, $\Psi$ and $E$ are the scalar perturbations
(with comma denoting ordinary spatial derivative,
and Latin indices denoting spatial coordinates, which are
raised or lowered with the unit matrix), and $h_{ij}$ is the
tensor fluctuation satisfying
${h_i}^i = 0$, ${h_{ij}}^{,i} = 0$. 
The relevant action for a minimally coupled inflaton
$\phi$ is
\begin{eqnarray}
\label{fullaction}
S = \int d^4 x {\sqrt{-g}} [ {m_{\rm pl}^2 \over 16 \pi} R- 
{1\over 2} \partial_\mu \phi \partial^\mu \phi - V(\phi)]
\end{eqnarray}
where $R$ is the Ricci scalar, and $V$ is some potential.
Decomposing $\phi$ into a homogeneous part $\phi_0$
and fluctuation $\delta \phi$, we can expand the
above action keeping all (metric and inflaton) perturbations 
up to second order. It cleanly separates into two
parts, containing only scalar ($S_S$) and 
only tensor fluctuations ($S_T$) respectively \cite{mukhanov}:
\begin{eqnarray}
S_S &=& \int {d^4 x \over 2} [(u')^2 - \partial_k u \partial^k u
+ [(\phi_0'/H)''/(\phi_0'/H)] u^2 ]
\end{eqnarray}
\begin{eqnarray}
S_T &=& \int {d^4 x \over 2} [ {P_{ij}}' {P^{ij}}'
- \partial_k P_{ij} \partial^k P^{ij} 
+ (a''/a) P_{ij} P^{ij} ]
\end{eqnarray}
where ${}'$ denotes derivative with respect to $\tau$, and
$H$ is the Hubble parameter. The scalar fluctuation $u$
and the tensor fluctuation $P_{ij}$ are defined as
\begin{eqnarray}
\label{uPdef}
u
\equiv a(\delta\phi + \phi_0' \Psi (aH)^{-1}) \quad , \quad P_{ij} \equiv 
\sqrt{m_{\rm pl.}^2 / 32 \pi} a h_{ij}
\end{eqnarray}
They are chosen so that the corresponding action
resembles that for a scalar field with variable mass, which
can be readily quantized. The quantity $u/a$ can be viewed
as a gauge invariant form of the inflaton 
fluctuation.

The scalar fluctuation obeys:
\begin{eqnarray}
\label{ueqt}
u'' - \nabla^2 u - {(\phi_0'/H)'' \over (\phi_0'/H)} u = 0
\end{eqnarray}
Expanding $u(\tau, {\bf x})$ in plane waves, we have
\begin{eqnarray}
u(\tau, {\bf x}) = \int d^3 k [ u_{\bf k} (\tau) a_{\bf k} e^{i {\bf k} \cdot
{\bf x}} + u_{\bf k} (\tau)^* a_{\bf k}^\dagger e^{-i {\bf k} \cdot
{\bf x}} ]
\end{eqnarray}
where $a_{\bf k}$ and $a_{\bf k}^\dagger$ are the annihilation and
creation operators. Bold faced symbols here denote
(3D) spatial vectors. Imposing the commutation relations
$[a_{\bf k_1}, a_{\bf k_2}^\dagger] = \delta^3 ({\bf k_1} - {\bf k_2})$
and $[u(\tau, {\bf x_1}), u(\tau, {\bf x_2})'] = 
i \delta^3 ({\bf x_1} - {\bf x_2})$
implies that the mode functions $u_{\bf k}$ obey
\begin{eqnarray}
\label{unorm}
u_{\bf k} {u_{\bf k}'}^* - {u_{\bf k}}^* u_{\bf k}' = i/(2\pi)^3
\end{eqnarray}

From eq. (\ref{ueqt}), it is clear that at early times, 
when $k^2 \gg (\phi_0'/H)''/(\phi_0'/H)$,  
$u_{\bf k}$ is oscillatory in behavior (here, $k = |{\bf k}|$).
The Bunch-Davies
vacuum corresponds to the choice of $u_{\bf k} (\tau) 
= (2\pi)^{-3/2} (2k)^{-1/2} e^{-ik\tau}$ at early times. The
associated operator $a_{\bf k}$ annihilates
the Bunch-Davies vacuum: $a_{\bf k} |0 \rangle_{\rm B.D.} = 0$. 
More generally, one can choose a different set of mode
functions and equivalently a different set of creation
and annihilation operators:
\begin{eqnarray}
\label{vb}
u(\tau, {\bf x}) = \int d^3 k [v_{\bf k} (\tau) b_{\bf k} e^{i {\bf k} \cdot
{\bf x}} + v_{\bf k} (\tau)^* b_{\bf k}^\dagger e^{-i {\bf k} \cdot {\bf x}}]
\\ \nonumber 
v_{\bf k} (\tau) = \alpha_{k}^{\rm S} u_{\bf k} (\tau) + \beta_{k}^{\rm S}
 u_{\bf k} (\tau)^* 
\, , \,
b_{\bf k} = {\alpha_{k}^{\rm S}}^* a_{\bf k} - 
{\beta_{k}^{\rm S}}^* a_{\bf k}^\dagger
\end{eqnarray}
The operator $b_{\bf k}$ annihilates a different vacuum
$b_{\bf k} |0 \rangle = 0$. We will evaluate all expectation
values below using this vacuum, noting that the Bunch-Davies
vacuum is recovered if $\alpha_{k}^{\rm S} = 1$ and 
$\beta_{k}^{\rm S} = 0$. Note that the analog of 
condition (\ref{unorm}) for $v_{\bf k}$ implies the coefficients obey
\begin{eqnarray}
\label{alphaSnorm}
|\alpha_{k}^{\rm S} |^2 - |\beta_{k}^{\rm S} |^2 = 1
\end{eqnarray}
Note that the above choice of the vacuum is not the most
general possible \cite{bd}. It does describe some of the recent
constructions in the literature \cite{iscap}, and is adopted here
for simplicity. It is straightforward to generalize our
calculation to a wider class of vacua. 

We should emphasize here that, if $\beta_k^{\rm S} \ne 0$,
the state $|0\rangle$ defined above is no longer the
ground state within the context of the low energy theory
(i.e. the usual field theory calculation). 
As we have explained in the introduction, this is a deliberate
choice to model the effects of short distance physics -- 
from the point of view of the late time evolution of wave-modes,
the only effects that short distance physics has are through
the definition of the initial state. We refer to this state
as the vacuum (or effective vacuum) 
here, but the reader should keep in mind this
is not the vacuum in the usual sense of being the ground
state in the low energy effective theory unless $\beta_k^{\rm S} = 0$. 

Eq. (\ref{ueqt}) tells us that at late times,
$v_{\bf k} (\tau) \rightarrow A_{k} \phi_0'/H$ where
$A_{\bf k}$ is some time independent function of ${\bf k}$. 
So, if we define the quantity $\zeta(\tau, {\bf x}) \equiv 
u(\tau, {\bf x}) (H/\phi_0')$, its 
two-point correlation
has the following late time asymptote
\begin{eqnarray}
\langle \zeta(\tau, {\bf x_1}) \zeta(\tau, {\bf x_2}) \rangle
= \int d^3 k |A_{k}|^2 e^{i {\bf k} \cdot ({\bf x_1 - x_2})}
\end{eqnarray}
The definition of $\zeta$ here agrees with that of \cite{bardeen}
in the long wavelength limit. It equals the intrinsic curvature
fluctuations on comoving hypersurfaces i.e. quantifying
the local departure from flatness. 
This variable is conveniently
frozen when the proper wavelength greatly exceeds the Hubble
radius. We define the power spectrum of curvature fluctuations
to be $P_\zeta ({k}) \equiv (2\pi)^3 |A_{k}|^2$ \cite{Pzeta}.
 
While eq. (\ref{ueqt}) can be solved exactly by expanding 
$\phi_0'/H$ in slow-roll parameters, 
it is sufficient for our purpose here to obtain $|A_k |^2$ by 
simply matching the early and late time solutions
(for the Fourier mode $v_{\bf k} (\tau)$) 
at Hubble radius crossing (i.e. $k=aH$). This yields
\begin{eqnarray}
\label{Pzeta}
P_\zeta ({k}) &=& {\gamma_{k}^{\rm S} \over 2k^3} 
\left[{a^2 H^4 \over {\phi_0'}^2}\right]_{\rm cross} = {8\pi \over m_{\rm pl.}^2} 
{\gamma_{k}^{\rm S} \over 2k^3}
\left[{H^2 \over 2\epsilon}\right]_{\rm cross}
\\ \nonumber 
\gamma_{k}^{\rm S} &\equiv &
|\alpha_{k}^{\rm S}|^2 + 
|\beta_{k}^{\rm S}|^2 + 2 {\,\rm Re \,}(\alpha_{k}^{\rm S}
{\beta_{k}^{\rm S}}^*)
\end{eqnarray}
where the subscript 'cross' refers to evaluation at
Hubble radius crossing, and 
$\epsilon$ is the first slow-roll parameter and is defined as
$\epsilon \equiv (m_{\rm pl.}^2 / 16 /\pi)
(V^{-1} dV/d\phi_0)^2 = 
- H' / (a H^2) = (8 \pi / m_{\rm pl.}^2) 
({\phi_0'}^2 / (2 a^2 H^2))$.
The factor of $\gamma_{k}^{\rm S}$ quantifies the
deviation from the standard prediction (corresponding
to $\gamma_{k}^{\rm S} = 1$) due to the choice of vacuum.

The tensor fluctuations can be derived in an analogous fashion.
They obey an equation of motion of the form:
\begin{eqnarray}
\label{Pijeqt}
P_{ij}'' - \nabla^2 P_{ij} - (a''/a) P_{ij} = 0
\end{eqnarray}
Expanding in Fourier modes gives
\begin{eqnarray}
P_{ij} (\tau, {\bf x}) &=& \int d^3 k \sum_{\lambda=1}^2 [
{v}_{\bf k} (\tau, \lambda) \epsilon_{ij} ({\bf k}, \lambda)
e^{i {\bf k} \cdot {\bf x}} {b}_{\bf k} (\lambda) \\ \nonumber 
&& + {v}_{\bf k} (\tau, \lambda)^* \epsilon_{ij} ({\bf k}, \lambda)^*
e^{-i {\bf k} \cdot {\bf x}} {b}_{\bf k} (\lambda)^\dagger ]
\end{eqnarray}
where $\lambda$ is the polarization index, and 
$\epsilon_{ij}({\bf k}, \lambda)$
is the polarization matrix which obeys 
${\epsilon_i}^i = 0$, $k^i \epsilon_{ij} = 0$ and
$\epsilon_{ij} ({\bf k}, \lambda) \epsilon^{ij} ({\bf k}, \lambda')^* 
= \delta_{\lambda, \lambda'}$. We have used $v_{\bf k} (\tau, \lambda)$
and $b_{\bf k} (\lambda)$ to denote the mode function and
annihilation operators for gravitons -- the extra argument $\lambda$
distinguishes them from the corresponding quantities for scalar
fluctuations. As before, we can in general choose 
$v_{\bf k} (\tau, \lambda)$ to have the following behavior
at early times:
\begin{eqnarray}
\label{vbT}
v_{\bf k} (\tau, \lambda) \rightarrow
\alpha_{\bf k}^{\rm T} (\lambda) {e^{-ik\tau} \over (2\pi)^{3/2} \sqrt{2k}}
+ \beta_{\bf k}^{\rm T} (\lambda) {e^{ik\tau} \over (2\pi)^{3/2} \sqrt{2k}}
\end{eqnarray}
\begin{eqnarray}
\label{alphaTnorm}
{\rm where} \quad 
|\alpha_{\bf k}^{\rm T} (\lambda)|^2 - |\beta_{\bf k}^{\rm T} (\lambda) |^2 = 1
\end{eqnarray}

Eq. (\ref{Pijeqt}) implies that $v_{\bf k} (\tau, \lambda)$ behaves
at late times as $v_{\bf k} (\tau, \lambda) \rightarrow
A_{k} (\lambda) a(\tau)$ where $A_k (\lambda)$ is some time independent
function of ${\bf k}$ and $\lambda$.
From the definition of $P_{ij}$ in eq. (\ref{uPdef}), it is
therefore clear that each Fourier mode of $h_{ij}$ approaches a constant 
at late times. In other words, the two point correlation function
of the tensor fluctuations has the following future asymptote:
\begin{eqnarray}
&& \langle h_{ij} (\tau, {\bf x_1}) h^{ij} (\tau, {\bf x_2})
\rangle 
\\ \nonumber =
&& {32\pi \over m_{\rm pl.}^2} \int d^3 k \sum_{\lambda=1}^2 |A_{k} 
(\lambda) |^2 e^{i{\bf k} \cdot
({\bf x_1 - x_2})}
\end{eqnarray}
Similar to the scalar case, we define the power spectrum
of tensor fluctuations to be $P_{h} (k) \equiv
(2\pi)^3 (32\pi / m_{\rm pl.}^2) \sum_\lambda |A_{k} 
(\lambda) |^2$. Matching the solutions for $v_{\bf k} (\tau, \lambda)$
at early and late times at Hubble radius crossing $k = aH$, we
obtain \cite{Phcite}
\begin{eqnarray}
\label{Ph}
&& P_h (k) = {\gamma_k^{\rm T} \over 2 k^3} {64 \pi \over m_{\rm pl.}^2} 
\left[ H^2 \right]_{\rm cross}
\\ \nonumber
\gamma_k^{\rm T} &\equiv&  {1\over 2} 
\sum_{\lambda=1}^2 [|\alpha_{k}^{\rm T}(\lambda) |^2 + 
|\beta_{k}^{\rm T} (\lambda) |^2 + 2 {\,\rm Re \,}(\alpha_{k}^{\rm T} (\lambda)
{\beta_{k}^{\rm T} (\lambda)}^*)]
\end{eqnarray}

Note that we allow here, for the sake of generality, the possibility that
$\gamma_k^{\rm T} \ne \gamma_k^{\rm S}$ i.e. the departure
from the Bunch-Davies vacuum can be different for tensor and scalar modes.
As we will see, even in the simplest case where
$ \gamma_k^{\rm T} = \gamma_k^{\rm S}$, the consistency relation
between scalar and tensor modes is generally broken.

\section{Discussion}
\label{discuss} 

Eq. (\ref{Pzeta}) and (\ref{Ph}) are the main results of the last section.
The effect of the vacuum choice is encapsulated in the two quantities
$\gamma_k^{\rm S}$ and $\gamma_k^{\rm T}$ defined in the two equations.
Note that the coefficients $\alpha_k^{\rm S}$, 
$\beta_k^{\rm S}$, $\alpha_k^{\rm T} (\lambda)$ and $\beta_k^{\rm T} (\lambda)$,
which control $\gamma_k^{\rm S}$ and $\gamma_k^{\rm T}$ and
define the vacuum, are not completely arbitrary because they
have to satisfy normalization conditions eq. (\ref{alphaSnorm}) \& 
(\ref{alphaTnorm}). 
Let us now compute the quantities of interest: the scalar and tensor
spectral indices and the tensor to scalar amplitude ratio.
For historical reasons, the scalar and tensor power spectra 
are often rescaled and expressed as
\begin{eqnarray}
A_{\rm S}^2 (k) \equiv {4 \over 25} {k^3 \over 2 \pi^2} P_\zeta (k) 
\equiv A_{\rm S}^2 (k_0) (k / k_0)^{n_{\rm S} - 1} \\ \nonumber 
A_{\rm T}^2 (k) \equiv {1 \over 100} {k^3 \over 2 \pi^2} P_h (k) 
\equiv A_{\rm T}^2 (k_0) (k / k_0)^{n_{\rm T}} 
\end{eqnarray}
The symbol
$k_0$ represents some fiducial wave number. 
Therefore, the tensor to scalar ratio is
\begin{eqnarray}
\label{ATAS}
A_{\rm T}^2 / A_{\rm S}^2 = \epsilon \gamma_k^{\rm T} / \gamma_k^{\rm S}
\end{eqnarray}
and the scalar and tensor spectral indices are
\begin{eqnarray}
\label{nTnS}
n_{\rm S} &=& 1 - 6 \epsilon + 2 \eta + (d{\rm ln \gamma_k^{\rm S}}/d {\rm ln} k)
\\ \nonumber 
n_{\rm T} &=& - 2\epsilon + (d{\rm ln \gamma_k^{\rm T}}/d {\rm ln} k)
\end{eqnarray}
where $\epsilon$ is the first slow-roll parameter defined
after eq. (\ref{Pzeta}), and $\eta$ is the second slow-roll parameter, $\eta \equiv
(m_{\rm pl.}^2 / 8 \pi) V^{-1} (d^2 V / d \phi_0^2)
= \epsilon - [(d^2 \phi_0 /dt^2) / (H d\phi_0/dt)]$, where $t$ is
proper time. The above can be derived by noting that in the slow-roll
approximation, $3 H d \phi_0 / dt \sim - dV / d\phi_0$. 
Eq. (\ref{ATAS}) \& (\ref{nTnS}) 
agree exactly with the standard results \cite{liddle} if 
$\gamma_k^{\rm S} = \gamma_k^{\rm T} = 1$.

Clearly, the usual consistency relation \cite{consistency},
\begin{equation}
\label{consistency}
A_{\rm T}^2 / A_{\rm S}^2 = - n_{\rm T} / 2 \, ,
\end{equation}
is broken in general
unless a special relationship between the coefficients
$\gamma_k^{\rm T}$, $\gamma_k^{\rm S}$ and $\epsilon$ is satisfied:
\begin{eqnarray}
\label{condition}
({\gamma_k^{\rm T} /\gamma_k^{\rm S}}) \epsilon = \epsilon - 
({d {\rm ln} \gamma_k^{\rm T} / d {\rm ln} k})/2
\end{eqnarray}
Even in the simplest case where $\gamma_k^{\rm T} = \gamma_k^{\rm S}$
(i.e. the modifications to the vacuum for gravitons and for scalar
fluctuations are identical), the consistency relation is
not satisfied unless $\gamma_k^{\rm T}$ is independent of $k$ as well
\cite{kofman}.

Of course, for arbitrary $\gamma_k^{\rm S}$ and $\gamma_k^{\rm T}$, the spectrum
can display strong deviations from the power-law form, in which case
the spectral index is a poor parametrization of the shape of
the power spectrum. However, as pointed out by \cite{tanaka},
one expects $\gamma_k^{\rm S}$ (and $\gamma_k^{\rm T}$) to be close
to unity so that the particle background does not dominate over
the potential energy of the inflaton i.e. the vacuum is 'close'
to the Bunch-Davies form. To be concrete, 
suppose the vacuum is such that $\alpha_k
\sim 1$, and $|\beta_k | \ll 1$ (eq. \ref{vb}, \ref{vbT}), where
the S and T labels have been dropped since the same argument
applies to both scalar and tensor fluctuations.
It can be shown that 
$\gamma_k = |\alpha_k + \beta_k |^2 = 1 + 2 |\beta_k |^2 + 2 {\, \rm Re \,} (\alpha_k \beta_k^*)
\sim 1 + 2 {\, \rm Re \,}
(\beta_k)$, where in the last equality we have kept terms up to first order, and
where we have used $|\alpha_k |^2 - |\beta_k |^2 = 1$ (eq. \ref{alphaSnorm}, \ref{alphaTnorm}).
Putting the above into eq. (\ref{Pzeta}) and (\ref{Ph}), it can be seen that
the scalar and tensor power spectra take the form of a weakly modulated power-law
\begin{eqnarray}
\label{modulation}
A_{\rm S}^2 (k) \propto [1+ f_{\rm S} (k)] k^{n_{\rm S} -1} \sim k^{n_{\rm S} (k) - 1}
\\ \nonumber
A_{\rm T}^2 (k) \propto [1+ f_{\rm T} (k)] k^{n_{\rm T}} \sim k^{n_{\rm T} (k)}
\end{eqnarray}
where $f_{\rm S} (k) = 2 {\, \rm Re \,}(\beta_k^{\rm S})$ and
$f_{\rm T} (k) = \sum_{\lambda=1}^2 {\, \rm Re \,} (\beta_k^{\rm T} (\lambda))$, 
both of which are small. The description of the spectrum in terms of
a running power law is sensible, especially if the variation of
$f_{\rm S} (k)$ and $f_{\rm T} (k)$ with $k$ is sufficiently slow.
As we have remarked before, even in the simplest case where $f_{\rm S} (k) = f_{\rm T} (k)$, 
the consistency condition is broken unless $f_{\rm T}$ is $k$-independent 
(eq. \ref{condition}).

The arguments of \cite{tanaka} show that $|\beta_k| \lsim H m_{\rm pl.} / m_{\rm c}^2$, where
$m_{\rm c}$ is the mass scale at which new physics kicks in. 
If $m_{\rm c} = m_{\rm pl.}$, $\beta_k$ can be safely ignored for inflation at e.g. 
GUT scale
\cite{asnote}.
However, it is quite possible $H < m_{\rm c} < m_{\rm pl.}$ in which case, 
$|\beta_k|$, though small, can conceivably be comparable to (or even larger than) 
the slow roll parameters $\epsilon, \eta$.
In this case, while the tensor to scalar ratio is essentially unchanged
i.e. $A_{\rm T}^2 / A_{\rm S}^2 \sim \epsilon$ (since $\epsilon$ is small, 
modifications due to the vacuum
are higher order; eq. \ref{ATAS}), the power spectrum shapes could be modified by
a non-negligible amount compared to the departure
from scale invariance due to $\epsilon, \eta$ (eq. \ref{nTnS}, \ref{modulation}). 
Since upcoming precision measurements promise to constrain the spectral index (at least
for scalar modes) to percent-level accuracy, vacuum effects (possibly as large as
$\sim H m_{\rm pl.}/m_{\rm c}^2$) should not be ignored a priori.

We end this discussion by noting that the consistency condition (eq. \ref{consistency}) 
can also be broken in multiple-field models of inflation, without
resorting to a novel vacuum.
However, even for 
multiple-field models (with the standard vacuum), the consistency condition is not 
arbitrarily broken, but weakens to an inequality\cite{multifieldcons}
$\left\vert n_{\rm T}\right\vert \ge 2 {A_{\rm T}^2 / A_{\rm S}^2}$. 
On the other hand, even this weakened consistency condition can be violated by a suitable
choice of the vacuum in the single-field model we have considered
i.e. by choosing $\epsilon \gamma_k^{\rm T} / \gamma_k^{\rm S} >
|\epsilon - d{\,\rm ln}\gamma_k^{\rm T} / d{\,\rm ln} k / 2 |$.

Ultimately, to 
make progress, we need a better understanding of
what forms $\gamma_k^{\rm T}$ and $\gamma_k^{\rm S}$ might take -- 
this would help us decide whether
resorting to multiple-field models or modifying the vacuum
(or even developing alternatives to inflation) 
is a more plausible option in the event that a violation of
the consistency relation (eq. \ref{consistency}) is indeed observed.
Perhaps the most important (and in retrospect not
surprising) lesson of our exercise 
is that the key predictions
of inflation for the scalar and tensor fluctuations depend
critically on the choice of the vacuum state. While the Bunch-Davies
vacuum is perhaps the most natural one, fundamentally it
is not obvious whether short-distance physics will indeed
pick out this vacuum. This is an important question which deserves
further investigation. Our main result is that a specific condition on the
vacuum (eq. \ref{condition}) has to be obeyed for the well-known
consistency relation between scalar and tensor modes to be respected.

\acknowledgments

We thank Robert Brandenberger, Richard Easther, Brian Greene, Alan Guth, Gary Shiu 
and Ed Witten for useful discussions, and Scott Dodelson for his excellent notes.
LH and WHK are supported respectively by the DOE Outstanding Junior Investigator Program, and 
the Columbia University Academic Quality Fund.


\begin{thebibliography}{xxx}

\bibitem{inflation}
         A. H. Guth, Phys. Rev. D {\bf 23}, 347 (1981);
         A.~D.~Linde, Phys.\ Lett. {\bf B108} 389 (1982);
         A.~Albrecht \& P.~J.~Steinhardt, Phys. Rev. Lett {\bf48}, 1220 (1982).

\bibitem{fluc}
	W. Press, Phys. Scr. {\bf 21}, 702 (1980);
	V. Lukash, Pisma Zh. Eksp. Teor. Fiz, {\bf 31}, 631 (1980);
	V. Mukhanov \& G. Chibisov, JETP Lett. {\bf 33}, 532 (1981);
	A. Starobinsky, Phys. Lett. {\bf 117B}, 175 (1982);
        S. W. Hawking, Phys. Lett. {\bf 115B}, 295 (1982);
        A. H. Guth \& S. Y. Pi, Phys. Rev. Lett. {\bf 49}, 1110 (1982);
        J. M. Bardeen, P. J. Steinhardt \&  M. S. Turner, Phys.
	Rev. D {\bf 28}, 679 (1983); L. F. Abbott \& M. B. Wise, Nucl. Phys.
	B {\bf 244}, 541 (1984).

\bibitem{transplanck}
        J.~Martin \& R.~H.~Brandenberger, Phys. Rev. {\bf D63}, 123501 (2001);
        R.~H.~Brandenberger \& J.~Martin, Mod. Phys. Lett. {\bf A16}, 999 (2001);
        J.~C.~Niemeyer, Phys. Rev. {\bf D63}, 123502 (2001);
        J.~C.~Niemeyer \& R.~Parentani, astro-ph/0101451;
        A. Kempf, astro-ph/0009209;
        A.~Kempf \& J.~C.~Niemeyer, astro-ph/0103255.

\bibitem{note}
	Most authors so far explore the effect of modifying
	short-distance physics on scalar fluctuations 
	by examining quantum
	fluctuations of a massless scalar field. The equation
	of motion obtained is actually closer to that for tensor
	modes than for scalar modes.

\bibitem{consistency}
	A. A. Starobinsky, Sov. Astr. Lett., {\bf 11}, 133 (1985);
        E. D. Stewart \& D. H. Lyth, Phys. Lett. {\bf 302B}, 171 (1993).

\bibitem{gravitywaves}
	M. Kamionkowski, A. Kosowsky \& A. Stebbins, Phys. Rev. Lett. {\bf 78}, 2058 (1997);
	U. Seljak \& M. Zaldarriaga, Phys. Rev. Lett. {\bf 78}, 2054 (1997).
	
\bibitem{tanaka}
        T.~Tanaka, astro-ph/0012431,
        A.~A.~Starobinsky, Pisma Zh. Eksp. Teor. Fiz. {\bf 73}, 415 (2001)

\bibitem{bunchdavies}
	G. W. Gibbons \& S. W. Hawking, Phys. Rev. D {\bf 15}, 2738 (1977);
	T. S. Bunch \& P. C. W. Davies, Proc. R. Soc. London A {\bf 360}, 117 (1978).

\bibitem{lidsey}
        J.~E. Lidsey, A.~R.~Liddle, E.~W.~Kolb, E.~J.~Copeland, T.~Barreiro, and M.~Abney, Rev. Mod. Phys. {\bf 69}, 373 (1997).

\bibitem{huang}
	See also J. C. Hwang, PRD {\bf 48}, 3544 (1993),
	Class. \& Quantum Gravity {\bf 15}, 1401 (1998), 
	who derived similar expressions to ours for
	power spectra under a general vacuum. 
	For the scalar mode, 
	we focus here on the gauge invariant curvature
 	fluctuation, which reduces to the inflaton fluctuation
	under the uniform curvature gauge adopted by the above
 	ref.

\bibitem{mukhanov}
	L. P. Grishchuk, Sov. Phys. JETP {\bf 40}, 409 (1974); 
	V. F. Mukhanov, Phys. Lett. B {\bf 218}, 17 (1989).

\bibitem{bd}
	N. Birrell \& P. C. W. Davies, Quantum Fields in Curved Space,
	Cambridge University Press (1982); see also
	J. Lesgourgues, D. Polarski \& A. A. 
	Starobinsky, Nucl. Phys. {\bf B497}, 479 (1997).

\bibitem{iscap}
	R.~Easther, B.~Greene, W.~H.~Kinney \& G.~Shiu, hep-th/0104102.

\bibitem{bardeen}
	J.~M.~Bardeen, Phys. Rev. D {\b 22}, 1882 (1980).

\bibitem{Pzeta}
	Our $P_\zeta$ here is equal to $2 \pi^2 P_{\cal R} (k) / k^3$ 
	in the notation of \cite{lidsey}. Our choice of factors of
	$2\pi$ conforms to usual practice in large scale structure studies.


\bibitem{Phcite}
	$P_h$ here is equal to $2 \pi^2 P_g (k) / k^3$ in the notation of
	\cite{lidsey}. 

\bibitem{liddle}
	A.~R.~Liddle \& D.~H.~Lyth, Phys. Rep. {\bf 231}, 1 (1993).
	Note that $\eta$ here follows the definition
	in this paper rather than \cite{lidsey}. 

\bibitem{kofman}
	See e.g. L. Kofman et al., in preparation (2001) for $\gamma_k^{\rm T} \ne 
	\gamma_k^{\rm S}$; 
	see also G. Shiu \& S.-H. H. Tye, hep-th/0106274.

\bibitem{asnote}
	Starobinsky \cite{tanaka} argued that applying the same argument {\it today}
	gives a much stronger constraint. See \cite{iscap} however for exceptions.

\bibitem{multifieldcons}
        D.~Polarski and A.~A.~Starobinsky, Phys. Lett. {\bf 356B}, 196 (1995);
        J.~Garc\'\i a-Bellido and D.~Wands, Phys. Rev. D {\bf 52}, 6 739 (1995);
        M.~Sasaki and E.~D.~Stewart, Prog. Theor. Phys. {\bf 95}, 71 (1996).

\end{thebibliography}
\end{document}